\documentclass[prb,twocolumn]{revtex4}
\begin{document}
\newcommand{\be}{\begin{equation}}
\newcommand{\ee}{\end{equation}}
\newcommand{\ba}{\begin{eqnarray}}
\newcommand{\ea}{\end{eqnarray}}
\newcommand{\vk}{{\bf k}}
\newcommand{\vq}{{\bf q}}
\newcommand{\vp}{{\bf p}}
\newcommand{\vx}{{\bf x}}
\title{ Optical conductivity of one-dimensional narrow-gap semiconductors}
\author{Hyun C. Lee}
\email{hyunlee@phys1.skku.ac.kr}
\affiliation{BK21 Physics Research Division and Institute of Basic Science,
 Department of Physics,\\
Sung Kyun Kwan University, Suwon, 440-746
Korea}
\date{\today}
\begin{abstract}
The optical conductivities of two one-dimensional narrow-gap semiconductors, anticrossing quantum Hall edge states and
carbon nanotubes, are studied using bosonization method.
A lowest order renormalization group analysis indicates that the bare band gap can be treated perturbatively
at high frequency/temperature.   At very low energy scale the optical conductivity is dominated by the excitonic
contribution, while at temperature higher than a crossover temperature the excitonic features are eliminated 
by thermal fluctuations. In case of carbon nanotubes the crossover temperature scale is estimated to be 300 K.
\end{abstract}
\maketitle
\section{Introduction}
One-dimensional (1D) narrow-gap semiconductors can be realized in  anti-crossing
quantum Hall edge states (AQHE) \cite{kang} and carbon nanotubes (CNT) \cite{nano}.  
The gaps in these systems are {\it single} particle gaps and {\it not} many body
gaps.  Theoretically, they provide the  unusual condition that the bare band gap $t$ is 
much {\it smaller} than the characteristic Coulomb energy scale $E_c$.  Moreover,
strong quantum fluctuations are present in these
systems, reflecting the 1D character.

In understanding the excitation spectra of semiconductors and insulators, the excitons
which are the bound state of electron and hole play very important roles.
In three dimensional semiconductors  excitons 
can be treated  successfully by solving 
the Bethe-Salpeter equation \cite{sham}.
However, in the strong coupling regime of 1D systems,  
the perturbative approaches
are not expected to be reliable due to large quantum
fluctuations.  
If  the Coulomb
scale is much larger than the gap one might naively expect that  exciton
instability\cite{halperinrice} would occur.
However, this simple picture neglects \textit{screening} which is expected to be large
due to the smallness  of the gap.
It is unclear whether a bound state of an "electron" and a 
"hole"
can exist in the presence of strong quantum fluctuations.

Bosonization provides  a natural framework for studying
1D narrow gap semiconductors. Bosonization method allows the (almost) exact treatment of strong Coulomb 
interaction, and it also  transforms the bare band gap term into the non-linear cosine potentials
leading to a \textit{non-integrable} sine-Gordon (sG) type model.\cite{mine}
It was argued that under certain conditions the excitons can exist even for the  Coulomb
interaction much larger than the bare gap, being accompanied by the \textit{enhanced} single particle gap.\cite{mine}

In this paper we study the optical conductivities of  AQHE\cite{kang,mine}
and CNT in semiconducting phase, assuming the \textit{absence} of exciton instability.
A simple lowest order renormalization group analysis  shows that the cosine term [bare band gap] can be treated perturbatively
at high frequency/temperature.  At low frequency/tempeterature the excitonic contributions dominate the optical conductivity.
The optical conductivity depends on the temperature strongly. Especially at temperature higher than a 
crossover temperature scale $T_{\textrm{cr}}$ all of the exciton features are completely eliminated by thermal
fluctuations. The crossover temperature scale is estimated to be about 300 K for CNT.

In connection with the present work we note the studies on the optical conductivities of Mott insulators.\cite{controzzi,essler}
In those studies the Mott insulator problem is mapped to the \textit{exactly solvable} sG model, and the optical conductivity
is calculated using the form-factor approach based on  the integrability and the perturbation with respect to the
conformal field theory. In particular,  the  behaviours near two-particle production threshold have been determined exactly.
Even if the physical origin of the Mott gap is very different from our band gap, 
the optical conductivities share many similar features, especially
in high frequency region.

This paper is organized as follows: In Sec. II we introduce the models.  In Sec. III we set up the formalisms 
for the computation of optical conductivity. In Sec. IV and V, the results for 
the optical conductivities of AQHE and CNT are presented, respectively. 
We close this paper in Sec. VI with summary.

\section{Models}
First consider the spinless fermion case realized in the spin-polarized AQHE.\cite{kang,mine}
The system is modeled by  the following Hamiltonian
\ba
\label{H1}
H&=&H_0+H_{{\rm coul}}+H_{t},\nonumber \\
H_0&=&v_F \int dx \Big[-i  \psi^\dag_{R }\partial_x \psi_{R }+
i \psi^\dag_{L } \partial_x \psi_{L } \Big] \nonumber \\
&=&
 \pi v_F \int dx  \Big[ \rho_{R }^2+ \rho_{L }^2 
\Big], \\
H_{{\rm coul}}&=& \int dx dy  \frac{V(x-y)}{2} \rho(x) \, \rho(y),\nonumber \\
H_{t}&=&-t \int dx \,\Big[ \psi_{R }^\dag(x) \psi_{L }(x)+ {\rm H.c} \Big],
\ea
where $\rho(x)=\rho_R(x)+\rho_L(y)$.
The operator $\psi_{R } (\psi_{L })$ is the right-moving (left-moving) 
edge \textit{ electron} operator.
$\rho_{R }=:\psi^\dag_{R } \psi_{R }:$ is the (normal-ordered) right-moving
edge electron density operator ($\rho_{L }$ is similarly defined).
$V(x)=\frac{e^2}{\epsilon}\frac{1}{
\sqrt{x^2+a^2}}$ is the Coulomb interaction.  $a$ is taken be  the shortest length scale of our problem.
The Coulomb matrix element is $V(k)=\frac{ 2 e^2}{\epsilon} K_0( a |k|) \sim \frac{2 e^2}{\epsilon} \ln \frac{1}{|k| a}$.
The tunneling between the right-moving and left-moving electrons is modeled by
$H_{t}$.  Note that a \textit{single particle} gap  opens up near the
Fermi points due to this tunneling term, and this provides the bare band gap.

The interacting electron systems can be bosonized in a standard way.\cite{voit,wen,convention}
The phase fields are defined by
\be
\rho_R+\rho_L=\frac{1}{\pi} \partial_x \theta, \quad
\rho_R-\rho_L=\frac{1}{\pi} \partial_x \phi.
\ee
The effective bosonized action in imaginary time reads
\ba
\label{edge}
S&=&\int dx d\tau \Big[ \frac{i}{\pi} \partial_\tau \theta  \partial_x \phi +
\frac{v_F}{2\pi} \big[ (\partial_x \theta)^2 + (\partial_x \phi)^2 \big] \Big] \nonumber \\
&+& \frac{1}{2\pi^2} \int dx dy d \tau \Big[ V(x-y) \partial_x \theta(x)   \partial_y \theta(y) \Big] \nonumber \\
&-&
\frac{t}{\pi a} \int dx d\tau \cos(2 \theta(x,\tau)). 
\ea
Integrating out the dual phase field $\phi$ we obtain
\ba
\label{action}
S&=&\frac{1}{2\pi}\,T \sum_\omega \int \frac{d k}{2\pi}\,\Big[
\frac{\omega^2}{v_F}+v_F k^2(1+ \frac{V(k)}{\pi v_F} ) \Big] \nonumber \\
&\times&\theta(i \omega,k) \theta(-i \omega,-k) \nonumber \\
&-&\frac{t}{\pi a} \int d x d \tau \cos[2 \theta(x,\tau)].
\ea
The above action looks very similar to the sG model,
except for
the momentum-dependent Coulomb interaction $V(k)$. If $V(k)$ were momentum
independent (local interaction in real space), the action would be exactly 
that of sG model. 

Second we consider a model for the CNT in semiconducting phase.
For CNT it is necessary to introduces two bands \cite{balent}.  Including also spin degrees of freedom we 
need four species of fermions  $\psi_{R/L,i=1,2, \sigma=\uparrow,\downarrow}$,
and equivalently four species of boson phase fields  $\theta_{i=1,2, \sigma=\uparrow,\downarrow},
\phi_{i=1,2, \sigma=\uparrow,\downarrow}$.
It is convenient to introduce the charge/spin bosons 
\be
\theta_{i,\rho/\sigma}=\frac{1}{\sqrt{2}} \big( \theta_{i \uparrow} \pm \theta_{i \downarrow} \big),\quad
\phi_{i,\rho/\sigma}=\frac{1}{\sqrt{2}} \big( \phi_{i \uparrow} \pm \phi_{i \downarrow} \big).
\ee
Introduce also the in-phase (+) and out-of-phase (-) bosons [$ \nu=\rho/\sigma$]
\be
\theta_{\nu,\pm}=\frac{1}{\sqrt{2}} \big( \theta_{1 \nu} \pm \theta_{2 \nu} \big),\quad
\phi_{\nu,\pm}=\frac{1}{\sqrt{2}} \big( \phi_{1 \nu} \pm \phi_{2 \nu} \big).
\ee
In particular, the total charge  and current density  in imaginary time are given by
\be
\rho=\frac{2}{\pi}\partial_x \theta_{\rho +},\quad
j=i \frac{2}{\pi}\partial_\tau \theta_{\rho +}.
\ee
Now the  Hamiltonian for CNT in semiconducting phase is given by
\ba
H&=&H_0+H_{\textrm{Coul}}+H_t, \nonumber \\
H_0&=&\frac{v_F}{2\pi}\,\sum_{ i=\pm, \nu=\rho/\sigma}\,\int dx \Big[ (\partial_x \theta_{ \nu i})^2+
(\partial_x \phi_{ \nu i})^2 \Big], \nonumber \\
H_{\textrm{Coul}}&=&\frac{1}{2} (\frac{2}{\pi})^2\int dx dy  V(x-y) 
 [\partial_x \theta_{\rho +}(x) ] [\partial_y \theta_{\rho +}(y) ], \nonumber \\
H_t&=&-\frac{t}{\pi a} \sum_{i=1,2, \alpha=\uparrow,\downarrow} \Big[ \psi^\dag_{R i \alpha} \psi_{L i \alpha} + \textrm{H.c} \Big],
\ea
where $H_t$ gives rise to the bare band gap $t$.
The action in imaginary time is given by [$\nu \pm=\sigma +, \sigma -, \rho -$]
\ba
\label{nano}
S&=&S_{\rho +}+S_{\rho -} + S_{\sigma +}+S_{\sigma -}+S_t, \nonumber \\
S_{\rho +}&=& T \sum_\omega \int \frac{d k}{2\pi}\,\frac{1}{2\pi} \Big[ \frac{\omega^2}{v_F}+v_F k^2(1+\frac{4 V(q)}{\pi v_F})
\Big] \nonumber \\
&\times&\theta_{\rho +}(i\omega,k)  \theta_{\rho +}(-i \omega,-k), \nonumber \\
S_{\nu \pm}&=&\int dx d\tau \,\frac{1}{2\pi}\,
\Big[\frac{1}{v_F}(\partial_\tau  \theta_{ \nu \pm})^2+ v_F (\partial_x \theta_{\nu \pm} )^2 \Big], \nonumber \\
S_t&=&-\frac{t}{\pi a}\sum_{i=1,2, \alpha=\uparrow,\downarrow} \int dx d\tau \cos[2 \theta_{i \alpha}(x,\tau)].
\ea
Note that the Coulomb interaction acts only on the total charge sector ($\rho +$).
The bare band gap term  $S_t$ of the action Eq. (\ref{nano}) implicitly assumes that the bosons $\theta_{i \alpha}$ to be expressed
in terms of charge/spin bosons $\theta_{\nu \pm}$.

\section{Optical conductivity: Formalism}
The real part of the optical conductivity  $\sigma(\omega, T)$ can be computed from the Kubo formula.
\ba
\chi^R(\omega,q)&=&-i \frac{1}{L } \int_{-\infty}^\infty  dx \int_0^\infty d t  \, e^{i (\omega + i \epsilon) t-i q x}\,
\nonumber \\
&\times& \langle    [  J(x,t), J(0,0) ] \rangle,\nonumber \\
\sigma(\omega > 0,T)&=&-\frac{\textrm{Im} \Big[ \chi^R(\omega, q = 0) \Big]}{\omega},
\ea
where the superscript $R$ denotes the retarded Green function and $L$ is the system size.
Practically it is convenient to compute the correlation function  $\chi^R$ 
in imaginary time, and then analytically continue into the real time.
The current operator for the action Eq. (\ref{edge}) of AQHE in imaginary time is given by $
J=- \frac{i}{ \pi}\,\partial_\tau \theta$.
The correlation function $\chi$ in imaginary time can be expressed as
\ba
\chi(x-x^\prime,\tau-\tau^\prime)&=&-\langle J(x,\tau)  J(x^\prime,\tau^\prime) \rangle  \nonumber  \\
&=&
\frac{1}{\pi^2} \langle \partial_\tau \theta(x,\tau)   \partial_{\tau^\prime} \theta(x^\prime,\tau^\prime) \rangle.
\ea
Let us first compute the optical conductivity of Eq. (\ref{edge}) with vanishing bare band gap [$H_t=0$]. 
\be
\sigma(\omega,q) \sim \frac{ \pi v_q}{2} \delta(\omega-v_q q),\quad v_q= v_F \sqrt{1+\alpha \ln \frac{1}{ qa}},
\ee
which corresponds to the ideal conductivity.\cite{voit}
The impurity pinning effect  would broaden the above delta function peak.

To assess the importance of the band gap term $H_t$ in perturbation theory  it is useful to  
investigate the renormalization group (R.G.) flow of
the coefficient of $H_t$, $\mu \equiv \frac{t}{\pi a}$ of Eq. (\ref{edge}, \ref{nano}).
In this paper we work out only the lowest order contributions to the R.G. flow. 
Let us first consider AQHE case Eq. (\ref{edge}).\cite{comment}
The boson field $\theta=\theta_s+\theta_f$ is split into the slow and fast part, and the fast part is integrated out.
\ba
S_{t, slow}&=&-\frac{\mu}{2}\, \int dx d\tau \langle e^{2 i (\theta_s+\theta_f)}+e^{-2 i (\theta_s+\theta_f)} \rangle_f \nonumber \\
&=&-\mu \int dx d\tau \cos(2 \theta_s)\,e^{-2 \langle \theta_f \theta_f \rangle_f },
\ea
where $\langle \cdots \rangle_f$ denotes the average over the fast degrees of freedom.
The average of $\theta_f$ requires the specification of the momentum-energy range. We choose to integrate over the
whole frequency and reduce the momentum cut-off step by step.[$\xi_k=\sqrt{1+ \frac{V(k)}{\pi v_F}}$]
\ba
 \langle \theta_f(0) \theta_f(0) \rangle_f&=&\int_{-\infty}^{\infty}\frac{d \omega}{2\pi}\,
\int_{\Lambda^\prime< |k| < \Lambda} \frac{dk}{2\pi} \nonumber \\
&\times& \frac{\pi}{\omega^2/v_F+v_F k^2 \xi_k^2}, \nonumber \\
&=& \int_{\Lambda^\prime}^{\Lambda} \frac{d k}{2} \frac{1}{k \xi_k},\quad
.
\ea
From the above we can read off the lowest order R.G. equation\cite{voit}
\be
\label{rge}
\frac{d \mu}{\mu}=-\frac{d \Lambda}{\Lambda}(2-\frac{1}{\xi_\Lambda}).
\ee
Integrating Eq. (\ref{rge}) we obtain [$\Lambda_1 > \Lambda_2$] the R.G. flow
\ba
\label{flow}
\frac{\mu_1}{\mu_2}&=&\left( \frac{\Lambda_2}{\Lambda_1} \right )^2 \nonumber \\
&\times& \exp\Big[-
\frac{2}{\alpha}( \sqrt{1+\alpha \ln \frac{1}{\Lambda_1 a}}- \sqrt{1+\alpha \ln \frac{1}{\Lambda_2 a}}) \Big].
\ea
Taking $\Lambda_1$ to be the bare momentum cut-off , we can set $\Lambda_1 a =1$.
Clearly the R.G. flow Eq. (\ref{flow}) tells us that the coupling constant $\mu$ becomes larger as the cut-off decreases,
and that eventually below  a certain   energy/momentum scale the perturbative expansion in $\mu$ fails.
Recalling the bare value of $\mu \sim \frac{t}{\pi a}$,  the perturbative calculation would break down below the momentum scale
$\Lambda_{\textrm{cr}}$, 
at which the renormalized $\mu$ becomes order of $\frac{E_c}{\pi a}$, where $E_c=\frac{e^2}{\epsilon a}$
 is the Coulomb energy scale. 
For strong Coulomb interaction,  the relation  $1 < \alpha \ln \frac{1}{\Lambda_{\textrm{cr}} a}$ is satisfied, and 
 $\Lambda_{\textrm{cr}}$ is given by
$\Lambda_{\textrm{cr}} \sim \frac{1}{a} \sqrt{\frac{ t}{E_c}}$. Then, the crossover energy scale is given by
\be
\label{crossover}
\omega_{\textrm{cr}}=T_{\textrm{cr}}=\frac{e^2}{\epsilon} \Lambda_{\textrm{cr}}=\sqrt{t E_c}.
\ee
The above crossover energy scale $\omega_{\textrm{cr}}$ coincides with the \textit{soliton} mass estimated by completely
different argument.\cite{mine}

At energy scale below $ \omega_{\textrm{cr}}$ the cosine band gap term cannot be treated perturbatively anymore. 
But at sufficiently low energy most of the excitations would occur near the bottom of cosine potential, and then we can expand
the cosine term in power series. 
This expansion can be shown to be valid by rescaling the boson field in the limit of strong Coulomb interaction.
\cite{mine}
\ba
\label{expand}
S&\sim& \frac{1}{2\pi} T \sum_\omega \int \frac{d k}{2\pi} \Big[\frac{\omega^2}{v_F}+v_F k^2 \xi_k^2+\frac{4 t}{a} \Big]
\nonumber \\
&\times&  \theta(-i\omega,k) \theta(i\omega,k) \nonumber \\
&-&\frac{2}{3} \frac{t}{\pi a} \int dx \tau : \theta^4(x,\tau):,
\ea
where $: :$ denotes the normal ordering.[Or equivalently, the exclusion of tadpole diagrams]
The computation of optical conductivity from Eq. (\ref{expand}) is rather straightforward.
The $\theta$ boson  of Eq. (\ref{expand}) describes the exciton degrees of freedom, and the exciton dispersion relation is
determined by the quadratic part of Eq. (\ref{expand}).\cite{mine}

Now let us explicitly work out the perturbative calculation of $\chi(x-x^\prime,\tau-\tau^\prime)$ in $\mu$.
\ba
\label{perturb}
\chi(1,2)&=& \frac{1}{\pi^2} \partial_{\tau_1} \partial_{\tau_2} \,
\frac{ \int D[\theta ] \,\theta(1) \theta(2) \,e^{-S} }{\int D[\theta ] e^{-S}} \nonumber \\
&=& \frac{1}{\pi^2}\lim_{\alpha , \beta \to 0} \, \frac{1}{i^2 \alpha \beta}\,  \nonumber \\
&\times& \partial_{\tau_1} \partial_{\tau_2} \,
\frac{ \int D[\theta ] e^{ i \alpha \theta(1) }e^{i \beta \theta(2)} e^{-S} }{\int D[\theta ] e^{-S} } .
\ea
The last expression of Eq. (\ref{perturb}) is convenient for the perturbative calculation in $\mu$. 
The disconnected parts of the numerator are cancelled by the denominator. In the second order of $\mu$ we find,
\ba
\chi(1,2) &\sim& \frac{4}{\pi^2}\, \frac{\mu^2}{2} \, \partial_{\tau_1} \partial_{\tau_2} \,\int d3 \,d4
\langle \theta(1) \theta (3) \rangle \cdot \langle \theta(4) \theta(2) \rangle \,\nonumber \\
&\times& \exp\big[-(\langle \theta(0) \theta(0) \rangle - \langle \theta(3) \theta(4) \rangle )],
\ea
where $\int d3 \equiv \int d x_3 d \tau_3$.
Here  the cosine term is assumed to be $\cos\theta$.
Define $M(3,4)=4 \mu^2  \exp\big[-(\langle \theta(0) \theta(0) \rangle - \langle \theta(3) \theta(4) \rangle )]$.
Then,
\be
\chi(1,2) \sim \frac{1}{\pi^2}\,\partial_{\tau_1} \partial_{\tau_2} \,\int d3 d4 \,
 \langle \theta(1) \theta (3) \rangle \, M(3,4)  \,\langle \theta(4) \theta(2) \rangle.
\ee 
After Fourier transform we get [$D(1,2)\equiv \langle \theta(1) \theta(2) \rangle$]
\ba
\label{sigma1}
\chi(i\omega,q)&=&\frac{1}{\pi^2}  (i \omega )(-i\omega) D(i \omega,q) M(i \omega,q) D(i \omega,q) \nonumber \\
&\sim& \frac{1}{\pi^2} \, \frac{\omega^2}{D^{-1}(i \omega,q)-M(i\omega,q)}.
\ea
In the last line of Eq. (\ref{sigma1}) we have used an approximation analogous to the Dyson summation. In this context,
$M(i\omega,q)$ plays a role of "self-energy".
The explicit expression for the optical conductivity which is valid for the frequency  higher than $\omega_{\textrm{cr}}$ is
\ba
\sigma(\omega)&=&[- \omega \textrm{Im} M^R(\omega,q=0)]/\Big[[\textrm{Im} M^R(\omega,q=0)]^2 \nonumber \\
&+&[(D^R)^{-1}(\omega,q=0)-\textrm{Re} M^R(\omega,q=0)]^2 \Big ],
\ea
where the superscript $R$ denotes the retarded Green function.

The "self-energy" $M(x \tau,0)$ can be re-expressed as  $M(x \tau,0)=e^{-F(x \tau)},\;\;\omega_k \equiv v_F |k| \xi_k$, where 
\be
F(x \tau)=T \sum_\omega \int \frac{d k}{2\pi} (1-e^{-i kx - i \omega \tau}) 
\frac{\pi v_F}{\omega^2+\omega_k^2}.
\ee

The CNT case can be worked out similarly, and only the results will be shown below.
The R.G. equation is given by
\be
\frac{d \mu}{\mu}=-\frac{d \Lambda}{\Lambda} \Big(\frac{5}{4}-\frac{1}{4} \xi_\Lambda \Big).
\ee
The R.G. flow and the crossover energy scale are given by
\ba
& &\frac{\mu_1}{\mu_2}=\left( \frac{\Lambda_2}{\Lambda_1} \right )^{5/4}\,\nonumber \\
&  &\times\exp\Big[ -
\frac{1}{2\alpha} \big( \sqrt{1+\alpha \ln \frac{1}{\Lambda_1 a}}-
 \sqrt{1+\alpha \ln \frac{1}{\Lambda_2 a}} \big ) \Big], \\
& &\omega_{\textrm{cnt},\textrm{cr}}=E_c (\frac{t}{E_c})^{4/5}.
\ea
The self-energy is given by
\ba
\label{cntcross}
M_{\textrm{cnt}}(x\tau)&=&\left( \frac{a}{\sqrt{x^2+v_F^2 \tau^2}} \right)^{3/2}\,e^{-F_{\textrm{cnt}}(x\tau)}, \nonumber \\
F_{\textrm{cnt}}(x\tau)&=&\frac{T}{4}  \sum_\omega \int \frac{d k}{2\pi} (1-e^{-i kx - i \omega \tau}) 
\frac{\pi v_F}{\omega^2+\omega_k^2}.
\ea
The charge part of the action for the excitions which is valid at very low energy is
\ba
\label{expand2}
S_{\rho +}&=&\frac{1}{2\pi} T \sum_\omega \int \frac{d k}{2\pi} \Big[\frac{\omega^2}{v_F}+v_F k^2 \xi_k^2+\frac{4 t}{a} \Big] 
\nonumber \\
&\times& \theta_{\rho +}(-i\omega,-k) \theta_{\rho +}(i\omega,k)+S_{\textrm{quartic}},
\ea
where $S_{\textrm{quartic}}$ is the quartic terms in 4 boson fields. Since we are not interested in the explicit calculations of
quantum corrections due to $S_{\textrm{quartic}}$ its detailed form is not displayed.
\section{The Optical conductivity of anticrossing quantum Hall edge states}
In this section we calculate the optical conductivity based on the action Eq. (\ref{edge}).
\subsection{ $T < T_{\textrm{cr}}$}
For simplicity, consider $T=0$ case.
When $\omega \gg \omega_{\textrm{cr}}$ the cosine band gap term can be treated perturbatively as discussed in Sec. III.
The  "self-energy" $M(i\omega,q)$ is given by
\ba
M(i\omega,q=0) &\sim& \mu^2\,\int d x \,d\tau \,e^{-\epsilon \sqrt{x^2+v_F^2 \tau^2} + i\omega \tau} \,\nonumber \\
&\times& e^{-\frac{4}{\sqrt{\alpha}}\,
[\ln\frac{\sqrt{x^2+ v_F^2 \tau^2}}{a}]^{1/2}} 
\nonumber \\
&\sim& \frac{\mu^2}{\omega^2} e^{-\frac{4}{\sqrt{\alpha}}\,\sqrt{\ln 1/\omega^2}},
\ea
where $\epsilon > 0 $ is an infinitesimal  convergence factor which is set to zero afterward, and 
the subleading logarithmic corrections were neglected.
In terms of the optical conductivity 
\be
\label{edge1}
\sigma(\omega, T=0 ) \sim \frac{\mu^2 v_F^2}{\omega^5} e^{-\frac{4}{\sqrt{\alpha}}\,\sqrt{\ln 1/\omega^2}}, \quad
\omega \gg  \omega_{\textrm{cr}}.
\ee
Notice that $e^{-\frac{4}{\sqrt{\alpha}}\,\sqrt{\ln 1/\omega^2}}$ decreases slower than any other power law dependence.
This is a characteristic of long range Coulomb interaction, and it has been studied by H. Schulz.\cite{schulz}
The result Eq. (\ref{edge1}) should be compared with the optical conductivity of Mott insulators.\cite{controzzi}
\be
\sigma_{\textrm{Mott}} \sim \frac{\mu^2 v_F^2}{\omega^5} \omega^{4 \beta^2}, \quad \omega \gg M_{\textrm{soliton}},
\ee
where $M_{\textrm{soliton}}$ is the soliton mass of sG model which is specified by
$S_{sG}=\int d^2 x \big[ \frac{1}{16 \pi} (\partial \theta)^2+2 \mu \cos (\beta \theta)\big ]$.
Formally, our result Eq. (\ref{edge1}) corresponds to the limit $\beta \to 0$, namely deep in the semiclassical limit.
For the acutal comparison with the experimental data we might need the R.G. improved perturbation theory\cite{giamarchi} as has
been done in the Mott insulator case.\cite{controzzi}
That implies that the coupling constant $\mu$ becomes running coupling constant $\mu(\omega)$, and  Eq. (\ref{edge1}) can
be re-expressed as
$\sigma(\omega, T=0 ) \sim \frac{\mu^2(\omega)}{\omega}$. $\mu(\omega)$ should be determined by solving the higher order
R.G. equations. In the lowest order R.G. the result Eq. (\ref{edge1}) is reproduced.

In the opposite case $\omega \ll \omega_{\textrm{cr}}$, the cosine term can be expanded into power series and the excitonic 
contribution will dominate the optical conductivity.
Calculating the optical conductivity using the expanded action Eq. (\ref{expand}) and neglecting the quantum correction
due to the quartic term  $\theta^4$ we obtain
\be
\label{excitonpeak} 
\sigma(\omega,T=0) \sim \frac{\pi v_F}{2} \delta(\omega-\sqrt{\frac{4 t v_F}{a}}),\quad  \omega \ll  \omega_{\textrm{cr}}.
\ee
The quantum corrections due to  the quartic term which introduce frequency dependent self-energy 
would broaden the sharp peak feature of Eq. (\ref{excitonpeak}).

The calculation of optical conductivity near $\omega \sim \omega_{\textrm{cr}}$ requires a non-perturbative treatment 
which is not available in our problem. But we can expect a peak at $\omega=2 \omega_{\textrm{cr}}$ corresponding to the
renormalized particle-hole (soliton-antisoliton) production.\cite{controzzi}
In analogy with the result obtained from  sG case \textit{away from} the Luther-Emery point( $\beta^2=1/2$),  
the square root singularity is not expected near the two particle threshold  in our case since
formally our results imply $\beta \to 0$ as discussed above.\cite{controzzi}
\subsection{$T > T_{\textrm{cr}}$}
In this high temperature regime, the band gap cosine term can be treated perturbatively over the \textit{whole} frequency range.
It is because $\max(\omega,T)$ cuts off the R.G. flow. 
In other words, when $\omega > T > T_{\textrm{cr}}=\omega_{\textrm{cr}}$, the frequency $\omega$ cuts off the R.G. flow, and then 
the R.G. flow  clearly lies in the perturbative regime $\omega > \omega_{\textrm{cr}}$.
When  $\omega <  T_{\textrm{cr}} < T$ the temperature $T$ cuts off the R.G. flow, and the R.G. flow  clearly lies in the 
perturbative regime $T >  T_{\textrm{cr}}$.
Thus in order to obtain the optical conductivity it suffices to compute the "self-energy" $M(i\omega,q)$.

At high frequency we can use the optical conductivity obtained for $T=0$ case.
\be
\label{edge2}
\sigma(\omega, T) \sim \frac{\mu^2 v_F^2}{\omega^5} e^{-\frac{4}{\sqrt{\alpha}}\,\sqrt{\ln 1/\omega^2}}, \quad
\omega > T \gg \omega_{\textrm{cr}}.
\ee

At low frequency the self-energy $M(i\omega, q)$ should be evaluated at finite temperature.
Carrying out the frequency summation [$M(x \tau)=e^{-F(x \tau)}$]
\ba
F(x \tau)&=&\int \frac{d k}{2\pi} \frac{2 \pi v_F}{ \omega_k} 
\Big[(1+2 n_B( \omega_k))  \nonumber \\
&-&e^{ -i k x}[ e^{-\omega_k |\tau |}
+2 \cosh ( \omega_k \tau) n_B(\omega_k)]\Big].
\ea
After Fourier transform we find [$ T > T_{\textrm{cr}} \gg \omega$]
\be
M(i\omega,q=0) \sim \frac{\mu^2  v_F }{T^2} e^{-\frac{4}{\sqrt{\alpha}}\,\sqrt{\ln 1/T^2}}
.\ee
In terms of optical conductivity [$ T > T_{\textrm{cr}} \gg \omega$]
\ba
\label{edge3}
\sigma(\omega,T) &\sim&  \big (\omega \frac{\mu^2  v_F }{T^2} e^{-\frac{4}{\sqrt{\alpha}}\,\sqrt{\ln 1/T^2}} \big )
\nonumber \\
&/&\Big[
[\frac{\mu^2  v_F }{T^2} e^{-\frac{4}{\sqrt{\alpha}}\,\sqrt{\ln 1/T^2}}]^2 \nonumber \\
&+&[\omega^2/v_F-\frac{\mu^2  v_F }{T^2} e^{-\frac{4}{\sqrt{\alpha}}\,\sqrt{\ln 1/T^2}}]^2
 \Big ].
\ea
In the high temperature regime there are no exciton peaks and the features of multi-particle productions. They are eliminated 
by thermal fluctuations.

\section{The optical conductivity of semiconducting CNT}
In this section we calculate the optical conductivity based on the action Eq. (\ref{nano}).
The calculations are essentially identical with those of AQHE. 
\subsection{$T < T_{\textrm{cnt}, \textrm{cr}}$}
Consider $T=0$ case.
The self-energy at high frequency is given by [$\omega \gg \omega_{\textrm{cnt},\textrm{cr}}$]
\be
\label{cnt1}
M(i\omega,q=0) \sim \mu^2 v_F (\omega)^{+3/2-2} e^{-\alpha \sqrt{ \ln 1/\omega^2}}.
\ee
The exponent $3/2$ is due to the contributions from the channels other than charge.[$\rho-, \sigma \pm$]
In terms of the optical conductivity
\be
\sigma(\omega,T=0) \sim \frac{\mu^2 v_F^2}{\omega^{7/2}}\,
e^{-\alpha \sqrt{ \ln 1/\omega^2}},\quad \omega \gg \omega_{\textrm{cnt},\textrm{cr}}.
\ee
The optical conductivity at low frequency    is  dominated by the excitonic contribution. Using the action
Eq. (\ref{expand2}) the excitonic contribution can be calculated easily.
\be
\sigma(\omega) \sim \frac{\pi v_F}{2} \delta(\omega-\sqrt{\frac{ 4 t v_F}{a}}), \quad  \omega \ll \omega_{\textrm{cnt},\textrm{cr}}.
\ee

\subsection{$T > T_{\textrm{cnt}, \textrm{cr}}$}
When $\omega \gg T >  T_{\textrm{cnt}, \textrm{cr}}$, the self-energy Eq. (\ref{cnt1}) can be used.
\be
\sigma(\omega,T) \sim \frac{\mu^2 v_F^2}{\omega^{7/2}}\,
e^{-\alpha \sqrt{ \ln 1/\omega^2}},\quad \omega \gg T >  \omega_{\textrm{cnt},\textrm{cr}}.
\ee
When  $\omega <  T_{\textrm{cnt}, \textrm{cr}} < T$ the self-energy should be evaluated at finte temperature.
\be
\label{cnt2}
M(i\omega,T) \sim \frac{\mu^2 v_F}{\sqrt{T}}\,e^{-\alpha \sqrt{\ln 1/T^2}}.
\ee
The optical conductivity becomes [$T > T_{\textrm{cnt},\textrm{cr}} \gg \omega$]
\ba
\label{cnt3}
\sigma(\omega,T) &\sim& ( \omega \frac{\mu^2  v_F }{\sqrt{T}} e^{-\frac{1}{\sqrt{\alpha}}\,\sqrt{\ln 1/T^2}}) \nonumber \\
&/&\Big[
(\frac{\mu^2  v_F }{\sqrt{T}} e^{-\frac{4}{\sqrt{\alpha}}\,\sqrt{\ln 1/T^2}})^2 \nonumber \\
&+& (\omega^2/v_F-\frac{\mu^2  v_F }{\sqrt{T}} e^{-\frac{1}{\sqrt{\alpha}}\,\sqrt{\ln 1/T^2}})^2
 \Big ].
\ea
As in the case of AQHE the excitonic contributions and the features of multi-particle production are eliminated by thermal
fluctuations.

The crossover temperature scale of the semiconducting CNT can be estimated as follows:
the semiconducting gap $t$ is the order of $10 \textrm{mev}$, and the Coulomb energy scale can be taken to 
be $1 \sim 2 \textrm{eV}$.\cite{nano} Then, using the expression for the crossover energy scale Eq. (\ref{cntcross}) we get
\be
T_{\textrm{cr}} \sim  200 - 300 \textrm{K}.
\ee
Therefore, a drastic change of optical spectra of semiconducting CNT is expected  around room temperature.
The experimental verification of the above change of optical conductivity would be most interesting.
\section{Summary}
The optical conductivites of two kinds of 1D narrow-gap semiconductors, anticrossing quantum Hall edges and 
semiconducting carbon nanotubes, are studied using bosonization method. 
A lowest order R.G. calculation indicates that the tunneling term which gives rise to the bare band gap can be
treated perturbatively for frequency/temperature higher than a crossover scale.  
The crossover scale can be identified with the soliton mass of the associated sine-Gordon model.
Below the crossover energy scale the optical conductivity is dominated by excitonic contribution characterized by a sharp peak.
In particular, for the temperature much higher than the crossover scale, the optical conductivity can be determined over the whole
frequency range by perturbative method. The excitonic features are found to be eliminated by thermal fluctuations in the high 
temperature regime.
The crossover temperature scale of the semiconducting CNT is estimated to be around 300 K.
\begin{acknowledgments}
This work was  supported by the Korea Science and Engineering
Foundation (KOSEF) through the grant No. 1999-2-11400-005-5, and by the 
Ministry of Education through Brain Korea 21 SNU-SKKU Program.
\end{acknowledgments}

\end{document}